\documentclass[11pt]{article}

\usepackage[table]{xcolor}
\usepackage[preprint]{acl}
\usepackage{times}
\usepackage{latexsym}
\usepackage[T1]{fontenc}
\usepackage[utf8]{inputenc}
\usepackage{microtype}
\usepackage{inconsolata}
\usepackage{amsmath,amssymb}
\usepackage{graphicx}
\usepackage{booktabs,longtable,tabularx,array}
\usepackage{ragged2e,makecell,multirow,pifont}

\newcolumntype{L}[1]{%
  >{\RaggedRight\arraybackslash
    \hyphenpenalty=10000
    \exhyphenpenalty=10000}m{#1}}

\newcolumntype{C}[1]{%
  >{\centering\arraybackslash}m{#1}}

\newcolumntype{Y}{%
  >{\RaggedRight\arraybackslash
    \hyphenpenalty=10000
    \exhyphenpenalty=10000}X}

\newcommand{\workcite}[2]{%
  \makecell[l]{%
    #1\\[-1.4pt]
    {\fontsize{6.9pt}{7.4pt}\selectfont #2}%
  }%
}

\newcommand{\rowsep}{%
  \arrayrulecolor{black!18}%
  \hline
  \arrayrulecolor{black}%
}

\newcommand{\cmark}{\ding{51}}
\newcommand{\xmark}{\ding{55}}
\newcommand{\pmark}{\(\circ\)}

\newcommand{\RepMetaCell}[3]{%
  \textbf{Task:} #1\par
  \textbf{Type:} #2\par
  \textbf{Label:} #3%
}

\newcommand{\RepSourceCell}[2]{%
  \textbf{Text:} #1\par
  \textbf{Evidence:} #2%
}

\newcommand{\RepFlipCell}[2]{%
  \textbf{Flip:} #1\par
  \textbf{Rationale:} #2%
}

\title{Small Cues, Big Consequences: Learning Pivotal Cues for Multimodal Meme Classification}

\author{
  Akshit Sharma \and Prashant W. Patil \\
  CVPR Lab, MFSDSAI \\
  Indian Institute of Technology Guwahati \\
  \texttt{akshitsharma.rs@gmail.com} \quad
  \texttt{pwpatil@iitg.ac.in}
}

\begin{document}
\maketitle
\begin{abstract}
Memes often derive their harmful, hateful, or sarcastic meaning from small but decisive visual, textual, or cross-modal cues. Existing multimodal classifiers can miss such evidence when relying mainly on global image-text representations. We introduce MemeCF, a cue-focused benchmark of 9,895 memes across harm, hate, and sarcasm, with annotations identifying the modality and rationale of the pivotal evidence. We also propose MemePIVOT, a local-global architecture for meme classification. MemePIVOT uses frozen CLIP features, unbalanced optimal transport to align words with image patches while allowing irrelevant evidence to remain unmatched, and an evidential fusion head to combine local grounding with global meme context under uncertainty. Experiments on HarMeme, PrideMM, and MemeCF show consistent gains over strong text-only, image-only, multimodal, and vision-language baselines. Cross-dataset and ablation results further show that explicit pivotal-evidence modeling improves robustness and contributes meaningfully beyond global multimodal representations. Our code and dataset are publicly available at \url{https://github.com/AkshitSharma1/MemePIVOT}
\end{abstract}

\section{Introduction}
An internet meme is rarely the sum of an image and a caption. Its meaning often emerges from the tension between visual content, overlaid text, cultural references, symbols, humor, and social context. Like a punchline in comedy, a small visual or textual element can invert the meaning of an entire meme. A single word may redirect the target of a joke, a background symbol may introduce hateful meaning, a gesture may signal harm, or a subtle mismatch between image and text may turn an otherwise neutral post into sarcasm. This makes memes compact but highly expressive multimodal artifacts whose interpretation depends not only on what is present, but on which details matter.

\begin{table*}[t]
\centering

{\fontsize{7.9pt}{8.7pt}\selectfont

\setlength{\tabcolsep}{1.5pt}
\renewcommand{\arraystretch}{1.00}
\setlength{\arrayrulewidth}{0.22pt}

\renewcommand{\tabularxcolumn}[1]{m{#1}}

\begin{tabularx}{\textwidth}{
@{}
L{0.180\textwidth}
L{0.235\textwidth}
C{0.060\textwidth}
C{0.070\textwidth}
C{0.065\textwidth}
C{0.065\textwidth}
Y
@{}
}

\toprule

\textbf{Work}
&
\makecell[l]{\textbf{Source /}\\[-1pt]\textbf{Context}}
&
\makecell{\textbf{Multi-}\\[-1pt]\textbf{task}}
&
\makecell{\textbf{Multi-}\\[-1pt]\textbf{domain}}
&
\makecell{\textbf{Cue /}\\[-1pt]\textbf{CF}}
&
\textbf{Size}
&
\textbf{Context / Tasks}
\\

\midrule

\workcite{DisinfoMeme}{\cite{qu2022disinfomeme}}
&
Reddit; COVID-19, BLM, Veganism
&
\xmark
&
\cmark
&
\xmark
&
1,170
&
Disinformation memes: COVID-19, BLM, Veganism
\\
\rowsep

\cite{tanaka2022learning}
&
Meme websites; general
&
\xmark
&
NR
&
\xmark
&
7,500
&
Humor in general memes
\\
\rowsep

\workcite{HMC}{\cite{kiela2020hateful}}
&
Self-generated memes
&
\xmark
&
NR
&
\pmark$^{*}$
&
10,000
&
Hateful meme detection
\\
\rowsep

\workcite{MultiOFF}{\cite{suryawanshi2020multioff}}
&
Social media; U.S. election
&
\xmark
&
\xmark
&
\xmark
&
743
&
Offensive meme detection; U.S. election
\\
\rowsep

\workcite{Harm-C}{\cite{pramanick2021detecting}}
&
Google Images; COVID-19
&
\pmark
&
\xmark
&
\xmark
&
3,544
&
Harmful COVID-19 memes and targets
\\
\rowsep

\workcite{Harm-P}{\cite{pramanick2021momenta}}
&
Google Images; U.S. politics
&
\pmark
&
\xmark
&
\xmark
&
3,552
&
Harmful U.S. politics memes and targets
\\
\rowsep

\workcite{CrisisHateMM}{\cite{bhandari2023crisishatemm}}
&
Twitter, FB, Reddit; war
&
\pmark
&
\xmark
&
\xmark
&
4,723
&
Russia--Ukraine war hate/target classification
\\
\rowsep

\workcite{PrideMM}{\cite{shah2024memeclip}}
&
FB, Twitter, Reddit; LGBTQ+
&
\cmark
&
\xmark
&
\xmark
&
5,063
&
LGBTQ+ movement: hate, target, stance, humor
\\
\rowsep

\workcite{MultiBully}{\cite{maity2022multitask}}
&
Twitter, Reddit; code-mixed
&
\cmark
&
\xmark
&
\xmark
&
5,854
&
Code-mixed cyberbullying: bully, sentiment, emotion, sarcasm, harmfulness
\\
\rowsep

\workcite{MMSD2.0}{\cite{qin2023mmsd2}}
&
Twitter; sarcasm memes
&
\xmark
&
NR
&
\pmark$^{**}$
&
$\sim$24.3K
&
Multimodal sarcasm detection
\\
\rowsep

\workcite{GuardHarMem}{\cite{elamrany2025guardharmem}}
&
Multiplatform harmful memes
&
\pmark
&
\pmark
&
\xmark
&
$\sim$16.6K
&
--
\\
\rowsep

\workcite{GOAT-Bench}{\cite{lin2026goatbench}}
&
Curated meme benchmark
&
\cmark
&
\cmark
&
\xmark
&
6K+
&
Hatefulness, misogyny, offensiveness, sarcasm, harmfulness
\\

\midrule

\textbf{MemeCF (Ours)}
&
\textbf{Curated multi-source benchmark}
&
\textbf{\cmark}
&
\textbf{\cmark}
&
\textbf{\cmark}
&
\textbf{9,895}
&
\textbf{Counterfactual meme understanding across hate, harm, and sarcasm}
\\

\bottomrule

\end{tabularx}
}

\caption{
Comparison of MemeCF with existing meme benchmarks.
CF = counterfactual-style supervision;
\cmark = yes; \xmark = no; \pmark = partial; NR = not clearly reported.
For cue/CF supervision,
\pmark$^{*}$ denotes confounder-based supervision and
\pmark$^{**}$ denotes debiased-cue supervision.
}
\label{tab:dataset_comparison}

\end{table*}
The growing influence of memes has made their automatic understanding an important problem for online safety and content moderation. Memes shape political discourse, reinforce social identities, mobilize communities, and influence public opinion. At the same time, they are increasingly used to spread hateful narratives, harmful stereotypes, targeted harassment, sarcasm-driven abuse, and other forms of online harm. Detecting such content is challenging because harmful meaning is often indirect, culturally grounded, and visually sparse. A meme may appear benign when viewed globally, while a small phrase, object, symbol, facial expression, or cross-modal relation determines whether it is hateful, harmful, sarcastic, or benign.

This sparse-evidence structure creates a central challenge for existing multimodal models. Many vision-language systems rely primarily on global image-text representations, which are effective for capturing broad semantic alignment but can overlook the fine-grained evidence that drives meme interpretation. In many cases, most of the image and text provide context, while only a small set of pivotal cues determines the final label. As a result, a model may predict the correct class for the wrong reason, rely on superficial correlations, or fail when the decisive cue is visually small, semantically subtle, or surrounded by distracting context. Reliable meme classification therefore requires more than recognizing the overall scene or reading the full caption; it requires identifying the pivotal visual, textual, and cross-modal evidence that supports the decision.

To study this problem, we introduce \textsc{MemeCF}, a cue-focused benchmark for fine-grained meme classification. \textsc{MemeCF} contains 9,895 memes annotated across three socially important subtasks: hate, harm, and sarcasm. These subtasks capture three semantically distinct forms of socially consequential interpretation: identity-directed hostility, threat or injury, and pragmatic incongruity. Unlike label-only meme datasets, \textsc{MemeCF} explicitly focuses on memes whose labels are driven by small but decisive textual, visual, or joint image-text details. Each meme is paired with cue-focused evidence annotations that describe the pivotal detail supporting the classification decision. 

Building on this cue-focused formulation, we propose MemePIVOT, an evidential local-global architecture for multimodal meme classification. MemePIVOT is designed around the observation that meme labels are often governed by a small set of pivotal cues. Rather than relying only on global image-text features, MemePIVOT explicitly learns local evidence by aligning textual tokens with image patches. It uses unbalanced optimal transport to identify fine-grained word--patch correspondences, constructs cross-modal evidence from these alignments, and encodes the resulting interactions with a local Transformer. In parallel, a global branch captures holistic image-text context. The local and global branches are then combined through an evidential fusion module that models confidence, uncertainty, and disagreement. This allows MemePIVOT to jointly learn about the overall meme context and the pivotal fine-grained evidence that determines the final prediction.

We evaluate MemePIVOT on three meme classification benchmarks: HarMeme, PrideMM, and our proposed MemeCF dataset. Across these datasets, MemePIVOT consistently improves over strong multimodal baselines and benchmark vision-language models. On HarMeme, MemePIVOT achieves 83.70 F1 / 92.92 AUROC; on PrideMM, it achieves 76.98 F1 / 84.05 AUROC; and on MemeCF, it achieves 76.88 F1 / 78.20 AUROC for Hate, 87.92 F1 / 94.33 AUROC for Harm, and 82.18 F1 / 85.16 AUROC for Sarcasm. Beyond standard in-dataset evaluation, we conduct cross-dataset experiments, compare against benchmark VLMs, and perform ablation studies to measure the impact of local evidence alignment, global context modeling, and evidential fusion. Our results show that modeling pivotal fine-grained evidence is critical for reliable meme classification, especially when the decisive cues occupy only a small portion of the image-text input.
\begin{figure*}[t]
    \centering
    \includegraphics[width=\linewidth]{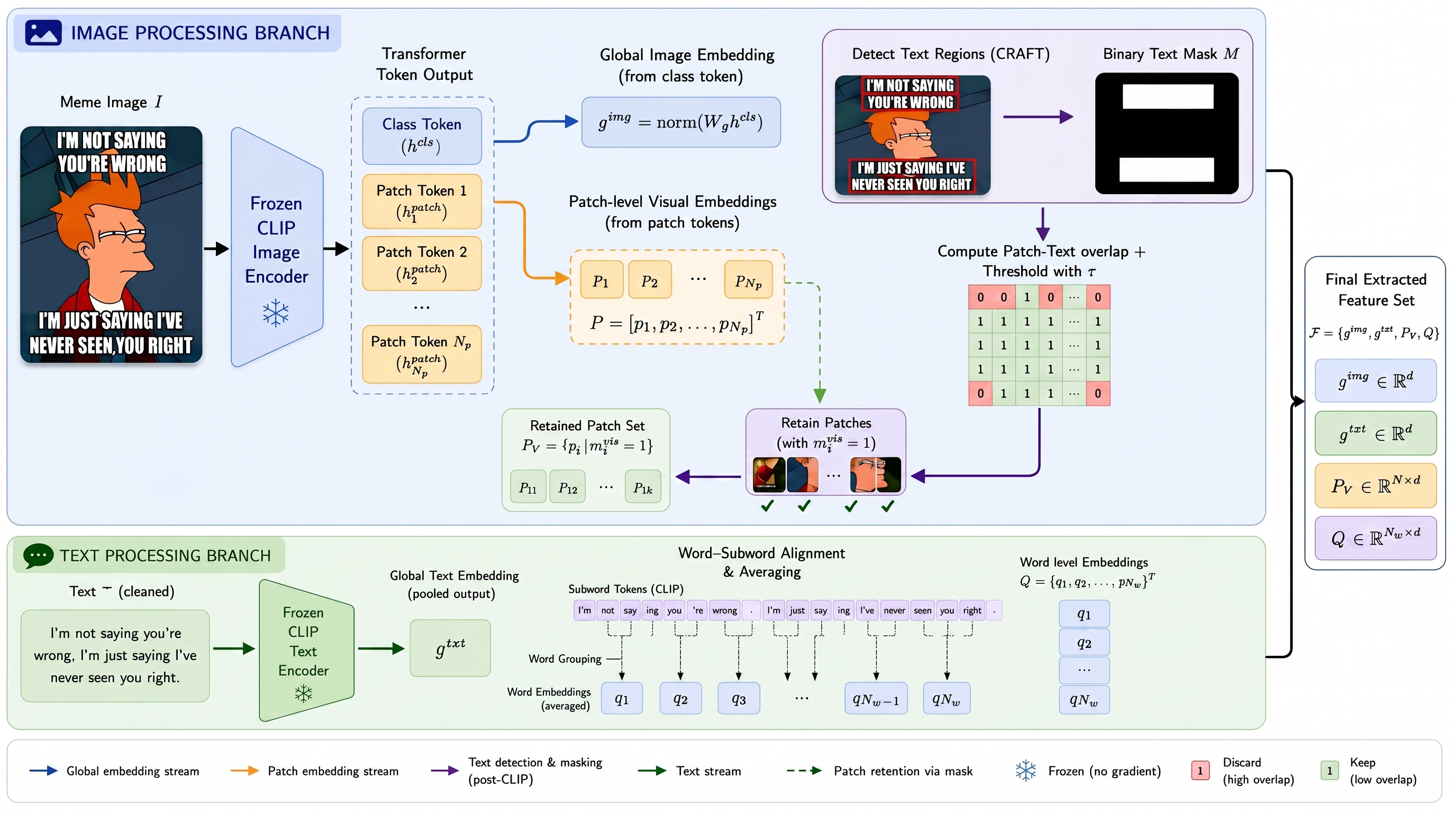}
\caption{Feature extraction step of MemePIVOT: frozen CLIP encoders produce global, patch-level, and word-level features with text-mask-guided patch retention.}
\label{fig:memepivot_feature_extraction}
\end{figure*}

\begin{figure*}[t]
    \centering
    \includegraphics[width=\linewidth]{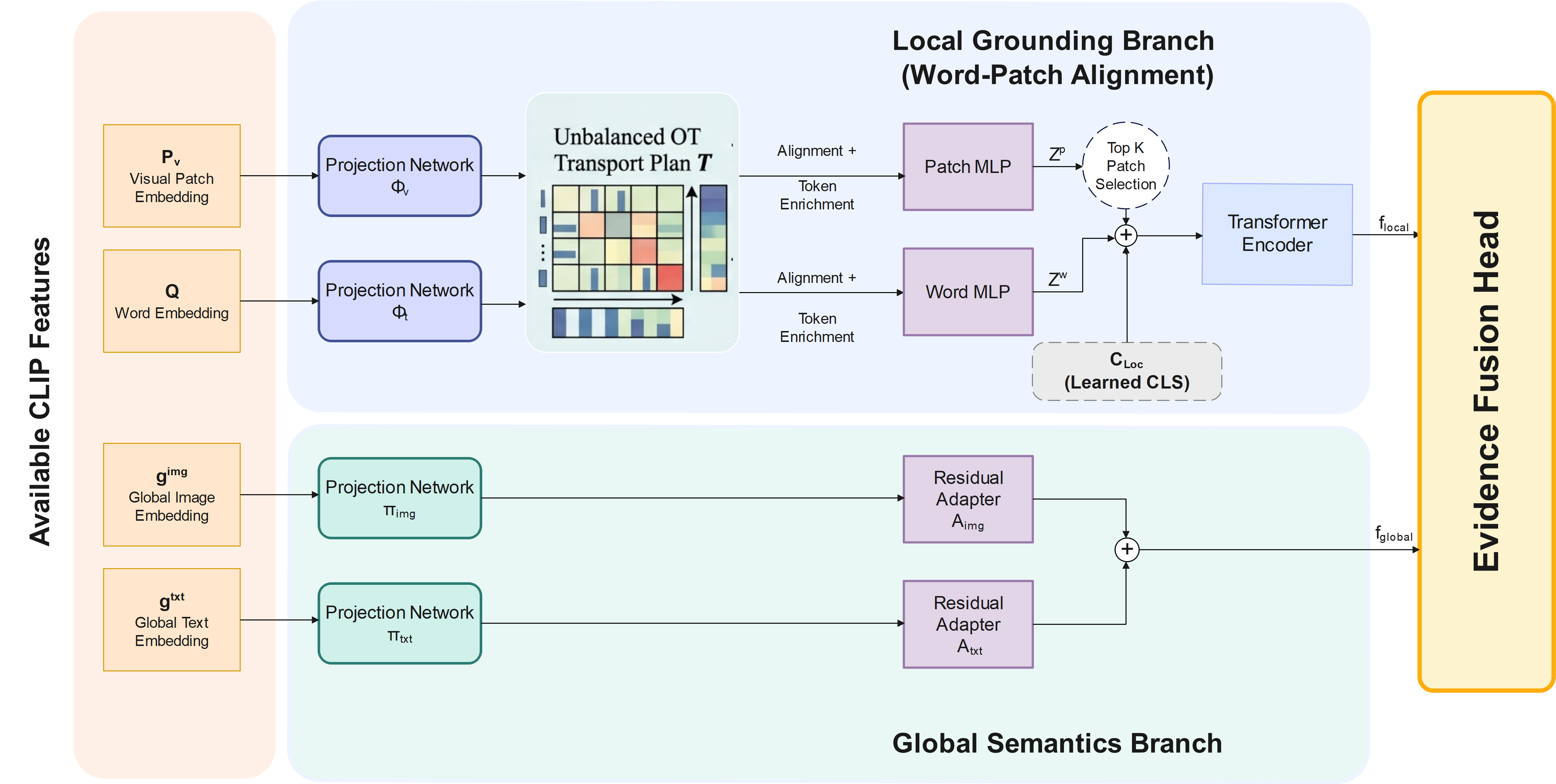}
    \caption{MemePIVOT architecture at a glance: extracted CLIP features are processed through local word--patch grounding and global semantic branches, then passed to the evidence-based fusion head for final classification.}
    \label{fig:memepivot_architecture}
\end{figure*}
\begin{figure*}[t]
    \centering
    \includegraphics[width=\linewidth]{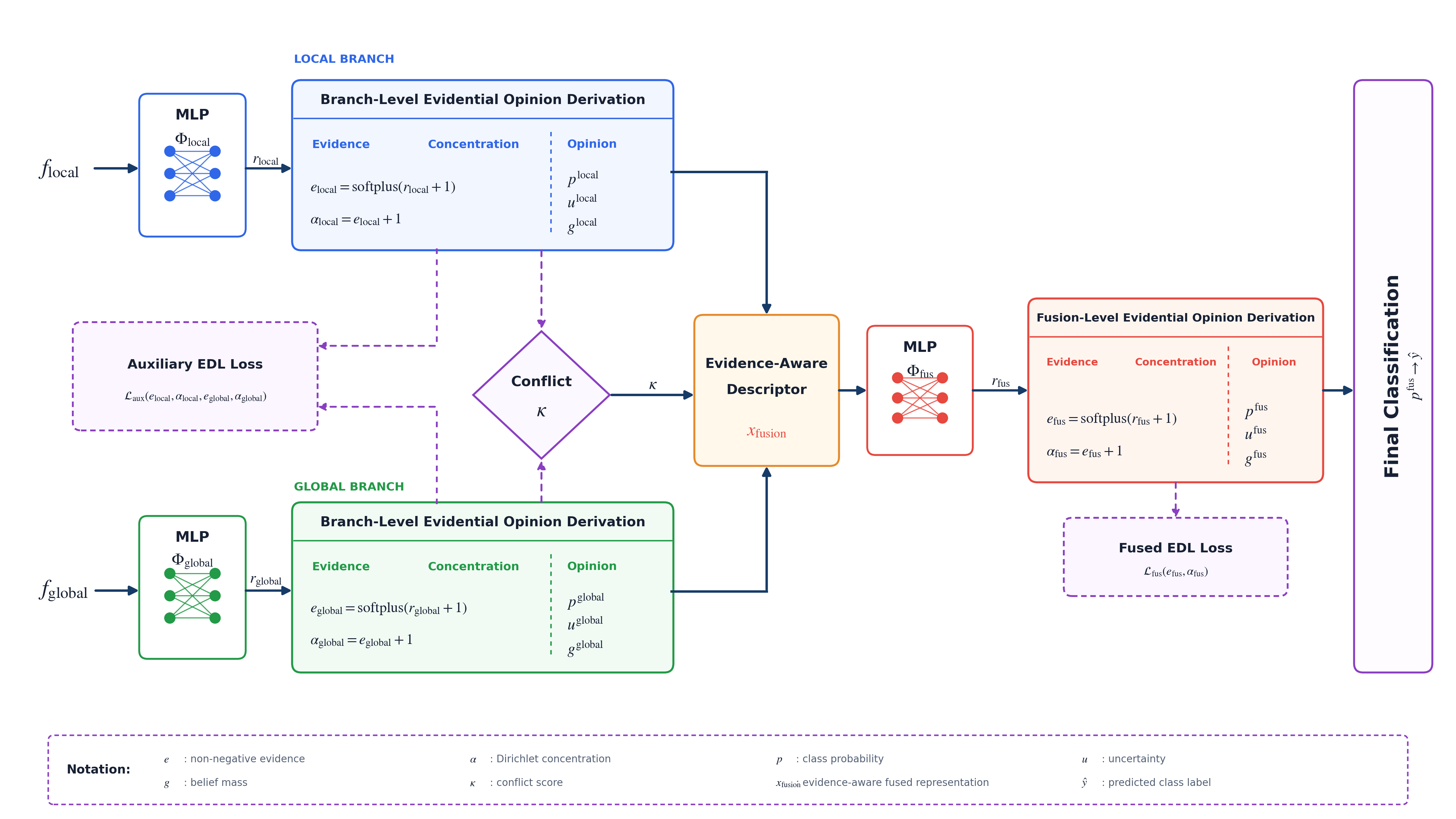}
    \caption{MemePIVOT evidence-based fusion head: local and global evidential opinions are conflict-aware fused to produce uncertainty-calibrated class predictions.}
    \label{fig:memepivot_evidential_fusion}
\end{figure*}
Our contributions are fourfold:
\paragraph{Contribution 1 --- Cue-focused benchmark.}
We introduce \textsc{MemeCF}, a cue-focused benchmark of 9,895 memes for fine-grained meme classification across hate, harm, and sarcasm. Unlike label-only meme datasets, \textsc{MemeCF} focuses on examples whose labels are driven by small but decisive textual, visual, or joint image-text details.

\paragraph{Contribution 2 --- Fine-grained evidence annotations.}
We provide cue-focused evidence annotations that describe the pivotal details supporting each classification decision. These annotations enable fine-grained dataset analysis and evidence-level evaluation beyond image-level labels, while classification models use only the meme image and extracted text as input.

\paragraph{Contribution 3 --- Pivotal evidence model.}
We propose \textsc{MemePIVOT}, an evidential local-global architecture for multimodal meme classification. \textsc{MemePIVOT} uses unbalanced optimal transport to align textual tokens with image patches, encodes the resulting local cross-modal interactions with a Transformer, and fuses local and global evidence under uncertainty.

\paragraph{Contribution 4 --- Comprehensive evaluation.}
We evaluate \textsc{MemePIVOT} across HarMeme, PrideMM, and \textsc{MemeCF}, including in-dataset performance, cross-dataset transfer, architectural ablations, and comparisons with benchmark vision-language models.
\section{Related Work}

\paragraph{Meme Benchmarks and Cue Evidence.}
Prior work has established meme classification as a challenging multimodal problem, with benchmarks spanning sentiment, emotion, offensiveness, hate, harm, cyberbullying, and sarcasm. \textsc{Memotion} and \textsc{MultiOFF} formalized meme analysis beyond ordinary image or text classification \citep{sharma2020semeval,suryawanshi2020multioff}, while \textsc{Hateful Memes} emphasized joint visual--textual reasoning through benign confounders \citep{kiela2020hateful}. Later datasets broadened the space to harmfulness, target identification, code-mixed cyberbullying, LGBTQ+ hate, fine-grained harm categories, and multimodal sarcasm \citep{pramanick2021detecting,maity2022multitask,shah2024memeclip,elamrany2025guardharmem,qin2023mmsd2,yue2024sarcnet}. In parallel, work on rationales and explanations has shown that final-label accuracy alone does not fully characterize model behavior: \textsc{ERASER} evaluates textual evidence spans \citep{deyoung2020eraser}, \textsc{VCR} studies vision-language rationales \citep{zellers2019vcr}, and meme-specific work has explored target-aware harmful meme detection and natural-language explanations \citep{pramanick2021detecting,sharma2022disarm,hee2023decoding}. These efforts provide important label-, target-, and rationale-level supervision, but they are not primarily organized around the specific setting where a small word, symbol, object, gesture, or image--text mismatch is sufficient to change the meme's interpretation.

\paragraph{Vision-Language Alignment and Uncertainty.}
Large vision-language representations such as CLIP have become standard backbones for multimodal meme understanding \citep{radford2021learning}. Meme-specific models adapt these representations through cross-modal feature interaction, lightweight CLIP-based modules, retrieval-guided contrastive learning, prompting, textual inversion, visually grounded context, or semantic tags \citep{kumar2022hateclipper,shah2024memeclip,mei2024retrieval,cao2022prompthate,burbi2023mapping,koushik2026trace,sharma2026memetag}. Beyond individual architectures, controlled studies have systematically examined encoder choice, fusion strategy, and modality reliance in multimodal meme classification \cite{Sharma_2026_CVPR}. While these approaches show the strength of pretrained vision-language features, global embeddings and generated summaries can blur the contribution of small local details when most of the meme provides context and only a small cue determines the label. Local cross-modal alignment offers a complementary direction: attention, graph reasoning, and feature interaction have been used to connect visual and textual information \citep{schifanella2016detecting,cai2019multi,liang2021multi,liang2022cmgcn}, while optimal transport provides a soft matching formulation across feature sets \citep{cuturi2013sinkhorn}. In text-embedded images, unbalanced optimal transport is particularly relevant because OCR tokens may be noisy, visual patches may be irrelevant, and not every word or region should be forced to align \citep{pham2020unbalanced}. Since local and global evidence may also disagree, evidential learning provides a way to represent both predictions and uncertainty \citep{sensoy2018evidential}. Together, these lines of work motivate meme classifiers that combine holistic image--text context with fine-grained local interactions and uncertainty-aware fusion.
\section{Methodology}

\subsection{Frozen CLIP Feature Extraction}

Given a meme image \(I\) and its text \(T\), we extract multimodal representations using frozen CLIP encoders. The CLIP vision transformer represents the image as one class token and \(N_p\) patch tokens. Instead of using only the default pooled image embedding, we retain the hidden token sequence before final pooling, \(H^{\mathrm{img}}=[h^{\mathrm{cls}};h^{\mathrm{patch}}_1;\dots;h^{\mathrm{patch}}_{N_p}]\). The class token gives the global image feature \(g_{\mathrm{img}}=\operatorname{norm}(W_vh^{\mathrm{cls}})\), while each patch token is projected into the CLIP embedding space as \(p_j=\operatorname{norm}(W_vh^{\mathrm{patch}}_j)\), forming \(P=[p_1,\dots,p_{N_p}]^\top\). Thus, \(g_{\mathrm{img}}\) preserves holistic visual semantics, whereas \(P\) provides local visual evidence from CLIP's internal patch-token states.

To reduce the effect of overlaid meme text on the local visual stream, we use CRAFT to detect text regions and rasterize them into a binary mask \(M\). This mask is not applied to the input image and therefore does not alter the frozen CLIP encoding. It is used only after feature extraction to measure text overlap for each CLIP patch. Let \(\Omega_j\) be the image region of the \(j\)-th patch. Its text coverage is \(\rho_j=|M\cap\Omega_j|/|\Omega_j|\), and the retention mask is \(m^{\mathrm{vis}}_j=\mathbf{1}[\rho_j\le\tau]\), where \(\tau\) is the text-overlap threshold. The retained local visual set is \(P_V=\{p_j\mid m^{\mathrm{vis}}_j=1\}\). This keeps the global image feature unchanged while allowing the local branch to ignore text-dominated patches.

For the text modality, the cleaned text \(T\) is encoded using the frozen CLIP text encoder. The pooled output is used as the global text embedding \(g_{\mathrm{txt}}\). Since CLIP uses subword tokenization, each word is aligned to its subword positions, and the corresponding projected hidden states are averaged to obtain word-level embeddings \(Q=[q_1,\dots,q_{N_w}]^\top\). The final feature set is \(\mathcal{F}=\{g_{\mathrm{img}},g_{\mathrm{txt}},P_V,Q\}\). The global branch uses \(g_{\mathrm{img}}\) and \(g_{\mathrm{txt}}\) for holistic meme semantics, while the local branch uses \(P_V\) and \(Q\) for fine-grained word--region grounding.

\subsection{Local--Global Multimodal Evidence Modeling}

\subsubsection{Local Grounding Branch}

The local branch models fine-grained alignment between words and visual regions. Given \(P_V\) and \(Q\), the visual and textual tokens are projected into a shared transport space of dimension \(d_{\mathrm{ot}}=384\) using the projection networks \(\Phi_v\) and \(\Phi_t\): \(\tilde p_j=\operatorname{norm}(\Phi_v(p_j))\) and \(\tilde q_i=\operatorname{norm}(\Phi_t(q_i))\). Their cross-modal cost is \(D_{ij}=1-\tilde q_i^\top\tilde p_j\), so semantically compatible word--patch pairs have lower cost.

Rather than forcing every word and patch to match, we solve an unbalanced optimal transport problem with dustbin states. This allows irrelevant words or patches to remain weakly matched, reducing spurious alignments. Let \(\mathbf{T}\) denote the resulting transport plan, where bold notation distinguishes it from the input text \(T\). For each word, the aligned visual summary is \(\hat p_i=\sum_j\mathbf{T}_{ij}\tilde p_j/(\sum_j\mathbf{T}_{ij}+\varepsilon)\), and for each patch, the aligned textual summary is \(\hat q_j=\sum_i\mathbf{T}_{ij}\tilde q_i/(\sum_i\mathbf{T}_{ij}+\varepsilon)\). The dustbin terms \(d_i^{(w)}\) and \(d_j^{(p)}\) measure unmatched word-side and patch-side evidence.

We form enriched local descriptors \(\bar z_i^{w}=[\tilde q_i;\hat p_i;\tilde q_i-\hat p_i;d_i^{(w)}]\) and \(\bar z_j^{p}=[\tilde p_j;\hat q_j;\tilde p_j-\hat q_j;d_j^{(p)}]\), each in \(\mathbb{R}^{3d_{\mathrm{ot}}+1}\). The Word MLP and Patch MLP map these descriptors into a local hidden dimension \(d_{\mathrm{local}}=256\), producing \(Z^w\) and \(Z^p\). To suppress noisy visual evidence, each patch is scored by its transported mass \(s_j=\sum_i\mathbf{T}_{ij}\), and only the top \(K\) patch tokens are retained as \(Z^p_{\mathrm{top}K}\). The word tokens, selected patch tokens, and learned local classification token \(C_{\mathrm{Loc}}\) are then concatenated and passed through a lightweight transformer: \(f_{\mathrm{local}}=\operatorname{Transformer}_{\mathrm{loc}}(C_{\mathrm{Loc}}\oplus Z^w\oplus Z^p_{\mathrm{top}K})_0\), where \(\oplus\) denotes token concatenation. This branch explicitly grounds text in relevant image regions, handles unmatched evidence through dustbins, and aggregates the most informative local cues.

\subsubsection{Global Semantic Branch}

The global branch preserves holistic meme-level semantics that may be lost when using only sparse local alignments. It operates on the frozen CLIP global embeddings \(g_{\mathrm{img}}\) and \(g_{\mathrm{txt}}\). The global projection networks \(\pi_{\mathrm{img}}\) and \(\pi_{\mathrm{txt}}\) map them into a shared latent dimension \(d_g=512\), giving \(u_{\mathrm{img}}=\pi_{\mathrm{img}}(g_{\mathrm{img}})\) and \(u_{\mathrm{txt}}=\pi_{\mathrm{txt}}(g_{\mathrm{txt}})\). These features are refined using residual adapters \(A(x)=x+W_2\sigma(W_1\operatorname{LN}(x))\), where \(\sigma(\cdot)\) is GELU and \(\operatorname{LN}(\cdot)\) is layer normalization. The adapted features are \(h_{\mathrm{img}}=A_{\mathrm{img}}(u_{\mathrm{img}})\) and \(h_{\mathrm{txt}}=A_{\mathrm{txt}}(u_{\mathrm{txt}})\), and the global representation is \(f_{\mathrm{global}}=h_{\mathrm{img}}\oplus h_{\mathrm{txt}}\in\mathbb{R}^{2d_g}\), where \(\oplus\) denotes feature concatenation. Unlike the local branch, this pathway does not perform patch pruning or token-level alignment; it retains broad visual context, overall text--image compatibility, and scene-level semantic cues.

Together, the two branches provide complementary evidence: \(f_{\mathrm{local}}\) captures grounded word--region interactions, while \(f_{\mathrm{global}}\) preserves high-level multimodal meaning.

\subsubsection{Evidence-Based Fusion Head}

After obtaining \(f_{\mathrm{local}}\) and \(f_{\mathrm{global}}\), the model fuses them using evidential opinions instead of directly averaging logits. For each branch \(m\in\{\mathrm{local},\mathrm{global}\}\), the branch MLP \(\Phi_m\) produces raw scores \(r_m=\Phi_m(f_m)\), which are converted into non-negative evidence \(e_m=\operatorname{softplus}(r_m)\) and Dirichlet parameters \(\alpha_m=e_m+\mathbf{1}\). Let \(N_c\) denote the number of classes and \(S_m=\sum_{c=1}^{N_c}\alpha_{m,c}\). The branch predictive probability, belief mass, and uncertainty are \(p_c^m=\alpha_{m,c}/S_m\), \(g_c^m=e_{m,c}/S_m\), and \(u^m=N_c/S_m\), respectively. Higher evidence therefore produces lower uncertainty, while weak evidence gives a flatter and less confident opinion.

The fusion module receives the evidence-aware descriptor 
\[
\begin{aligned}
x_{\mathrm{fusion}}
= [&\, e_{\mathrm{local}}; e_{\mathrm{global}}; g^{\mathrm{local}};
g^{\mathrm{global}}; \\ u^{\mathrm{local}};
&\, u^{\mathrm{global}}; \kappa \,].
\end{aligned}
\]
where the conflict term \(\kappa=(p_{+}^{\mathrm{local}}-p_{+}^{\mathrm{global}})^2\) measures disagreement between the local and global positive-class probabilities. This helps the model distinguish confident agreement from confident disagreement.

The fusion MLP \(\Phi_{\mathrm{fus}}\) maps \(x_{\mathrm{fusion}}\) to raw fused scores \(r_{\mathrm{fus}}=\Phi_{\mathrm{fus}}(x_{\mathrm{fusion}})\), which are converted into fused evidence \(e_{\mathrm{fus}}=\operatorname{softplus}(r_{\mathrm{fus}})\) and fused Dirichlet parameters \(\alpha_{\mathrm{fus}}=e_{\mathrm{fus}}+\mathbf{1}\). With \(S_{\mathrm{fus}}=\sum_{c=1}^{N_c}\alpha_{\mathrm{fus},c}\), the final predictive probability, belief mass, and uncertainty are \(p_c^{\mathrm{fus}}=\alpha_{\mathrm{fus},c}/S_{\mathrm{fus}}\), \(g_c^{\mathrm{fus}}=e_{\mathrm{fus},c}/S_{\mathrm{fus}}\), and \(u^{\mathrm{fus}}=N_c/S_{\mathrm{fus}}\). The final prediction is \(\hat y=\arg\max_c p_c^{\mathrm{fus}}\).

This fusion head acts as an evidence-aware arbitration module. It can rely more on the local branch when word--region grounding is confident, rely more on the global branch when holistic meme semantics are clearer, or reduce confidence when the branches are uncertain or contradictory.
The model is trained by supervising the fused output and both branch-level opinions. Let
\[
\begin{aligned}
\mathcal{L}_f 
&= \mathcal{L}_{\text{EDL}}(e_{\text{fus}}, \alpha_{\text{fus}}, y), \\
\mathcal{L}_{\text{loc}} 
&= \mathcal{L}_{\text{EDL}}(e_{\text{local}}, \alpha_{\text{local}}, y), \\
\mathcal{L}_{\text{glob}} 
&= \mathcal{L}_{\text{EDL}}(e_{\text{global}}, \alpha_{\text{global}}, y).
\end{aligned}
\]
The total loss is
\[
\mathcal{L}
=
\mathcal{L}_f
+
\lambda_{\text{local}}\mathcal{L}_{\text{loc}}
+
\lambda_{\text{global}}\mathcal{L}_{\text{glob}}
+
\lambda_{\text{uot}}\mathcal{L}_{\text{uot}}
+
\lambda_{\text{dust}}\mathcal{L}_{\text{dust}}.
\]

The auxiliary branch losses keep both local and global representations discriminative, while the transport and dustbin losses encourage stable alignment and controlled unmatched evidence. Overall, the final decision is formed by reconciling fine-grained grounding cues with holistic multimodal context on a per-sample basis.
\section{Experimental Details and Result Analysis}
\subsection{MemeCF Dataset}
\begin{table}[t]
\centering
\small
\setlength{\tabcolsep}{3pt}
\begin{tabular*}{\columnwidth}{@{\extracolsep{\fill}}lrrrr@{}}
\toprule
\textbf{Task} & \textbf{All 0/1} & \textbf{Train 0/1} & \textbf{Val 0/1} & \textbf{Test 0/1} \\
\midrule
Harm    & 1397 / 1621 & 1045 / 1218 & 139 / 163 & 213 / 240 \\
Hate    & 1500 / 1789 & 1127 / 1340 & 152 / 177 & 221 / 272 \\
Sarcasm & 1639 / 1949 & 1233 / 1458 & 164 / 195 & 242 / 296 \\
\bottomrule
\end{tabular*}
\caption{MemeCF statistics over 9,895 memes. Label 0/1 denotes negative/positive classes for each task: Harm, Hate, and Sarcasm.}
\label{tab:memecf_stats}
\end{table}

We introduce \textbf{MemeCF}, a counterfactual meme dataset for three binary tasks: harm, hate, and sarcasm. It contains 9,895 memes sourced from MultiBully, MMSD 2.0, PrideMM, HMC, HarMeme, and GuardHarMem~\citep{maity2022multitask,qin2023mmsd2,shah2024memeclip,kiela2020hateful,pramanick2021detecting,elamrany2025guardharmem}. Each instance includes an image, meme text, task label, modality-level counterfactual evidence indicating whether the label-changing cue lies in the text, image, or both, and supporting reasoning. Candidate counterfactuals were filtered through a LLM-as-judge protocol followed by human review. Further details, including image-selection criteria, examples, and extended distributions, are provided in the Supplementary Material.

Table~\ref{tab:memecf_stats} reports the task-wise label distribution. Each task is split into approximately 75\%/10\%/15\% train, validation, and test partitions while preserving label balance. 
\subsection{Dataset Configuration}
We evaluate MemePIVOT on three binary multimodal meme benchmarks: HarMeme, PrideMM, and MemeCF, covering harmful, hateful, and sarcasm-related meme understanding. Each dataset consists of image--text meme pairs \((I,T_o)\) with task-specific labels.

HarMeme~\citep{pramanick2021detecting} contains 3,544 COVID-19 memes labeled harmful or harmless and is evaluated using its official splits. PrideMM~\citep{shah2024memeclip} includes 5,063 LGBTQ+ memes, split into 4,328/228/507 train, validation, and test examples. MemeCF, our proposed counterfactual meme dataset, contains 9,895 memes across harm, hate, and sarcasm, with each task split into approximately 75\%/10\%/15\% train, validation, and test partitions while preserving label distributions, as shown in Table~\ref{tab:memecf_stats}. Models are trained on the training split, tuned on validation data, and evaluated on the held-out test split. Additional implementation details are provided in the Supplementary Material.

\subsection{Experimental Results}

\begin{table*}[t]
\centering
\scriptsize
\caption{Performance comparison across HarMeme and PrideMM using mean $\pm$ standard deviation. Single-run VLM results marked by $^\dagger$.}
\label{tab:harmeme_pridemm_grouped_results_seed101_42_256}
\resizebox{\textwidth}{!}{
\begin{tabular}{lccc ccc}
\toprule
\multirow{2}{*}{\textbf{Model}} 
& \multicolumn{3}{c}{\textbf{HarMeme}} 
& \multicolumn{3}{c}{\textbf{PrideMM}} \\
\cmidrule(lr){2-4} \cmidrule(lr){5-7}
& \textbf{Acc.} & \textbf{AUROC} & \textbf{F1}
& \textbf{Acc.} & \textbf{AUROC} & \textbf{F1} \\
\midrule

\multicolumn{7}{l}{\textit{Text-only models}} \\
\midrule
BERT~\cite{devlin2019bert}
& $75.83 \pm 1.25$ & $82.57 \pm 1.06$ & $68.31 \pm 0.86$
& $67.85 \pm 1.41$ & $78.02 \pm 0.59$ & $71.58 \pm 1.01$ \\

CLIPTextOnly~\cite{radford2021learning}
& $75.81 \pm 1.43$ & $80.01 \pm 1.58$ & $59.26 \pm 3.62$
& $69.87 \pm 2.10$ & $75.86 \pm 0.73$ & $69.86 \pm 1.13$ \\

\midrule
\multicolumn{7}{l}{\textit{Image-only models}} \\
\midrule
CLIPImgOnly~\cite{radford2021learning}
& $74.16 \pm 1.97$ & $82.92 \pm 0.83$ & $67.98 \pm 2.43$
& $62.65 \pm 1.93$ & $68.73 \pm 2.01$ & $60.01 \pm 3.47$ \\

\midrule
\multicolumn{7}{l}{\textit{Vision-language models}} \\
\midrule
LLaVA-1.5-7B$^{\dagger}$~\cite{liu2024improved}
& $72.88$ & $66.31$ & $43.53$
& $59.96$ & $45.27$ & $55.19$ \\

Qwen2.5-VL$^{\dagger}$~\cite{qwen25vl2025}
& $83.33$ & $90.13$ & $78.39$
& $75.12$ & $82.21$ & $75.09$ \\

\midrule
\multicolumn{7}{l}{\textit{Other multimodal models}} \\
\midrule
CLIPAdapter~\cite{gao2024clipadapter}
& $81.93 \pm 2.12$ & $91.14 \pm 0.71$ & $78.03 \pm 1.34$
& $74.89 \pm 1.20$ & $\mathbf{84.07 \pm 0.53}$ & $74.92 \pm 1.23$ \\

M2CLIP\_SEmoNet~\cite{dhiman2026enhancing}
& $81.87 \pm 0.92$ & $89.91 \pm 1.53$ & $73.33 \pm 2.08$
& $74.27 \pm 1.58$  & $82.11 \pm 1.01$  & $75.17 \pm 1.60$ \\

DORA~\cite{hossain2024deciphering}
& $82.81 \pm 1.39$ & $88.02 \pm 1.84$ & $76.01 \pm 2.15$
& $73.76 \pm 1.29$ & $81.74 \pm 0.53$ & $73.11 \pm 1.82$ \\
HateCLIPPER~\cite{kumar2022hateclipper}
& $85.73 \pm 0.61$ & $90.27 \pm 0.21$ & $80.84 \pm 0.60$
& $74.81 \pm 0.61$ & $82.54 \pm 0.49$ & $74.39 \pm 0.47$ \\

HarMDetect~\cite{elamrany2025guardharmem}
& $84.01 \pm 0.25$ & $90.24 \pm 1.27$ & $78.21 \pm 1.02$
& $73.51 \pm 0.12$ & $81.02 \pm 0.20$ & $73.15 \pm 0.53$ \\

MemeCLIP~\cite{shah2024memeclip}
& $84.75 \pm 0.28$ & $91.31 \pm 0.35$ & $81.70 \pm 0.74$
& $75.11 \pm 0.41$ & $82.92 \pm 0.53$ & $73.67 \pm 0.60$ \\

\textbf{MemePIVOT}
& $\mathbf{88.04 \pm 0.59}$ & $\mathbf{92.92 \pm 0.40}$ & $\mathbf{83.70 \pm 0.48}$
& $\mathbf{76.90 \pm 0.83}$ & $84.05 \pm 0.11$ & $\mathbf{76.98 \pm 0.73}$ \\
\bottomrule
\end{tabular}
}
\end{table*}

\begin{table*}
\centering
\footnotesize
\setlength{\tabcolsep}{2.5pt}
\renewcommand{\arraystretch}{1.08}
\caption{Performance comparison across MemeCF tasks. Multi-seed results are reported as mean $\pm$ standard deviation; single-run zero-shot VLM baselines are marked by $^\dagger$.}
\label{tab:memecf_grouped_results_seed101_42_256}
\resizebox{\textwidth}{!}{
\begin{tabular}{lccc|ccc|ccc}
\toprule
\multirow{2}{*}{\textbf{Model}} 
& \multicolumn{3}{c|}{\textbf{MemeCF-Hate}} 
& \multicolumn{3}{c|}{\textbf{MemeCF-Harm}} 
& \multicolumn{3}{c}{\textbf{MemeCF-Sarcasm}} \\
\cmidrule(lr){2-4} \cmidrule(lr){5-7} \cmidrule(lr){8-10}
& \textbf{Acc.} & \textbf{AUROC} & \textbf{F1}
& \textbf{Acc.} & \textbf{AUROC} & \textbf{F1}
& \textbf{Acc.} & \textbf{AUROC} & \textbf{F1} \\
\midrule

\multicolumn{10}{l}{\textit{Text-only models}} \\
\midrule
BERT~\cite{devlin2019bert}
& $63.08 \pm 3.26$ & $69.44 \pm 2.13$ & $69.78 \pm 3.36$
& $73.14 \pm 0.13$ & $81.34 \pm 0.14$ & $76.82 \pm 0.31$
& $68.59 \pm 2.81$ & $74.68 \pm 3.14$ & $73.72 \pm 1.73$ \\

CLIPTextOnly~\cite{radford2021learning}
& $59.43 \pm 3.96$ & $63.12 \pm 8.37$ & $71.08 \pm 0.89$
& $62.33 \pm 3.17$ & $72.51 \pm 0.79$ & $71.63 \pm 0.36$
& $64.13 \pm 5.70$ & $70.71 \pm 0.48$ & $74.52 \pm 2.04$ \\

\midrule
\multicolumn{10}{l}{\textit{Image-only models}} \\
\midrule
CLIPImgOnly~\cite{radford2021learning}
& $58.62 \pm 2.82$ & $62.85 \pm 2.24$ & $70.32 \pm 1.08$
& $71.30 \pm 3.11$ & $79.57 \pm 1.07$ & $77.30 \pm 1.10$
& $64.62 \pm 2.45$ & $69.53 \pm 1.98$ & $72.10 \pm 2.86$ \\
\midrule

\multicolumn{10}{l}{\textit{VLM baselines}} \\
\midrule
Gemma 3 Vision 12B$^{\dagger}$~\cite{gemma3team2025}
& $70.39$ & $75.32$ & $74.30$
& $62.69$ & $62.81$ & $69.77$
& $61.15$ & $67.07$ & $73.17$ \\

LLaVA-1.5-7B$^{\dagger}$~\cite{liu2024improved}
& $58.62$ & $68.90$ & $52.56$
& $53.64$ & $52.65$ & $59.92$
& $57.62$ & $58.67$ & $71.21$ \\

Qwen3-VL-8B$^{\dagger}$~\cite{qwen3vl2025}
& $68.76$ & $75.10$ & $74.84$
& $60.26$ & $64.87$ & $68.31$
& $63.75$ & $69.37$ & $74.77$ \\

Phi-4 Multimodal$^{\dagger}$~\cite{abouelenin2025phi4mini}
& $66.73$ & $73.52$ & $69.74$
& $58.50$ & $62.26$ & $59.48$
& $66.91$ & $72.10$ & $72.53$ \\
\midrule

\multicolumn{10}{l}{\textit{Other multimodal models}} \\
\midrule
CLIPAdapter~\cite{gao2024clipadapter}
& $68.29 \pm 1.25$ & $74.06 \pm 0.75$ & $73.25 \pm 1.71$
& $82.27 \pm 1.00$ & $90.11 \pm 0.83$ & $83.79 \pm 0.88$
& $77.20 \pm 1.67$ & $83.92 \pm 0.85$ & $80.45 \pm 2.72$ \\

M2CLIP\_SEmoNet~\cite{dhiman2026enhancing}
& $68.36 \pm 2.34$ & $74.97 \pm 1.12$ & $72.74 \pm 1.00$
& $84.91 \pm 0.48$ & $92.93 \pm 0.29$ & $86.13 \pm 0.40$
& $71.50 \pm 0.09$ & $77.85 \pm 0.98$ & $75.91 \pm 1.15$ \\

HarMDetect~\cite{elamrany2025guardharmem}
& $68.22 \pm 1.84$ & $72.57 \pm 1.11$ & $73.51 \pm 1.38$
& $83.91 \pm 1.24$ & $92.73 \pm 0.75$  & $85.31 \pm 1.31$
& $74.78 \pm 0.76$ & $82.00 \pm 0.15$ & $78.15 \pm 0.63$ \\

DORA~\cite{hossain2024deciphering}
& $65.86 \pm 1.15$ & $71.42 \pm 1.29$ & $73.12 \pm 0.57$
& $82.78 \pm 1.81$ & $91.90 \pm 1.79$ & $84.67 \pm 1.47$
& $73.36 \pm 0.65$ & $79.92 \pm 0.57$ & $77.55 \pm 1.68$ \\

MemeCLIP~\cite{shah2024memeclip}
& $66.60 \pm 0.23$ & $74.29 \pm 0.64$ & $66.10 \pm 0.10$
& $83.37 \pm 0.67$ & $92.61 \pm 0.23$ & $83.27 \pm 0.67$
& $76.83 \pm 0.88$ & $\mathbf{85.61 \pm 0.31}$ & $76.08 \pm 0.97$ \\

HateCLIPPER~\cite{kumar2022hateclipper}
& $71.08 \pm 0.62$ & $76.94 \pm 0.62$ & $75.52 \pm 1.38$
& $84.91 \pm 0.21$ & $92.98 \pm 0.12$ & $86.09 \pm 0.23$
& $77.06 \pm 1.09$ & $81.22 \pm 0.96$ & $80.57 \pm 0.95$ \\

ISSUES~\cite{burbi2023mapping}
& $70.59 \pm 1.05$ & $76.54 \pm 0.30$ & $74.47 \pm 1.39$
& $82.49 \pm 0.92$ & $90.93 \pm 1.20$ & $83.85 \pm 0.80$
& $77.20 \pm 1.55$ & $85.05 \pm 0.83$ & $80.52 \pm 1.47$ \\

\textbf{MemePIVOT} & $\mathbf{71.38 \pm 0.23}$ & $\mathbf{78.20 \pm 0.57}$ & $\mathbf{76.88 \pm 1.02}$
& $\mathbf{86.92 \pm 1.29}$ & $\mathbf{94.33 \pm 0.22}$ & $\mathbf{87.92 \pm 1.22}$
& $\mathbf{79.18 \pm 0.66}$ & $85.16 \pm 0.89$ & $\mathbf{82.18 \pm 0.71}$ \\
\bottomrule
\end{tabular}
}
\end{table*}
As shown in Tables~\ref{tab:harmeme_pridemm_grouped_results_seed101_42_256} and~\ref{tab:memecf_grouped_results_seed101_42_256}, MemePIVOT achieves the best performance on five of six HarMeme/PrideMM metrics and eight of nine MemeCF metrics. On HarMeme, it improves over the strongest non-MemePIVOT result by $2.31$ Acc., $1.61$ AUROC, and $2.00$ F1 points. On PrideMM, it improves Acc. and F1 by $1.78$ and $1.81$ points, respectively, while its AUROC of $84.05$ is $0.02$ points below CLIPAdapter's $84.07$. On MemeCF, MemePIVOT leads all three Hate metrics, all three Harm metrics, and Sarcasm Acc. and F1; MemeCLIP obtains the best Sarcasm AUROC. These results show strong and consistent performance across datasets, with the remaining difference confined to one metric.

\subsubsection{Cross-Dataset Evaluation}
\begin{table}[t]
\centering
\tiny
\caption{Cross-dataset hate-classification evaluation under dataset shift, with models trained on PrideMM and evaluated on HMC using the same task definition.}
\label{tab:pridemm_hmc_results}
\begin{tabular}{lccc}
\toprule
\textbf{Model} & \textbf{Acc.} & \textbf{AUROC} & \textbf{ F1} \\
\midrule
ISSUES~\cite{burbi2023mapping}           & 0.5796 & 0.6021 & 0.5938 \\
MemeCLIP~\cite{shah2024memeclip}         & 0.5776 & 0.5994 & 0.5879 \\
HarMDetect~\cite{elamrany2025guardharmem} & 0.5455 & 0.5861 & 0.5709 \\
MemePIVOT                                & \textbf{0.6106} & \textbf{0.6287} & \textbf{0.6358} \\
\bottomrule
\end{tabular}
\end{table}
We evaluate MemePIVOT alongside ISSUES, MemeCLIP, and HarMDetect under a controlled cross-dataset setting. Each model is trained on the PrideMM~\citep{shah2024memeclip} hate task and evaluated on the HMC~\citep{kiela2020hateful} test set, preserving the same binary hate-classification task definition while changing the dataset distribution. As shown in Table~\ref{tab:pridemm_hmc_results}, MemePIVOT achieves the strongest performance across all three metrics, outperforming the strongest comparison baseline, ISSUES, by $3.10$ Acc., $2.66$ AUROC, and $4.20$ F1 percentage points. This indicates that the learned pivotal-evidence representation transfers when evaluated under cross-dataset/out-of-domain setting

\section{Ablation Studies}
In this section, we present an ablation study evaluating the contribution of each proposed model component to overall performance. In addition, the supplementary material includes an ablation study benchmarking state-of-the-art LLMs on meme classification using the HarMeme test set. The results highlight two key findings: (i) trained meme classification models continue to outperform state-of-the-art LLMs as of 2026, and (ii) state-of-the-art LLMs are substantially more expensive to deploy, making their large-scale use for internet-scale meme moderation infeasible.
\subsection{Impact of each Component on Overall Model Performance}
\begin{table}[t]
\centering
\tiny
\setlength{\tabcolsep}{3pt}
\caption{Cumulative ablation study on HarMeme showcasing impact of each component. 
G denotes the global branch, L denotes the local branch, E denotes the evidential head, and OT denotes the optimal-transport loss. Components are added progressively from top to bottom. Results are reported as mean $\pm$ standard deviation.}
\label{tab:harmeme_ablation}
\begin{tabular}{ccccccc}

\toprule
\textbf{G} & \textbf{L} & \textbf{E} & \textbf{OT} 
& \textbf{Acc.} & \textbf{AUROC} & \textbf{F1} \\
\midrule
$\checkmark$ & -- & -- & -- 
& $84.56 \pm 0.38$ 
& $91.38 \pm 0.98$ 
& $78.12 \pm 1.84$ \\

$\checkmark$ & $\checkmark$ & -- & -- 
& $85.86 \pm 1.12$ 
& $91.10 \pm 0.82$ 
& $79.48 \pm 0.54$ \\

$\checkmark$ & $\checkmark$ & $\checkmark$ & -- 
& $87.23 \pm 0.71$ 
& $92.08 \pm 0.43$ 
& $81.52 \pm 1.18$ \\

$\checkmark$ & $\checkmark$ & $\checkmark$ & $\checkmark$ 
& $\mathbf{88.04 \pm 0.59}$ 
& $\mathbf{92.92 \pm 0.40}$ 
& $\mathbf{83.70 \pm 0.48}$ \\
\bottomrule
\end{tabular}
\vspace{-2mm}
\end{table}

We perform a cumulative ablation to isolate the effect of each component on model performance using the HarMeme dataset, as shown in Table~\ref{tab:harmeme_ablation}. Adding the local branch raises Acc. from $84.56$ to $85.86$ and F1 from $78.12$ to $79.48$, while AUROC changes from $91.38$ to $91.10$. Adding the evidential head further improves Acc., AUROC, and F1 to $87.23$, $92.08$, and $81.52$, respectively. Adding the optimal-transport loss produces the strongest configuration, reaching $88.04$ Acc., $92.92$ AUROC, and $83.70$ F1, corresponding to gains of $0.81$, $0.84$, and $2.18$ points over the preceding configuration. Overall, the full model improves over the global-only variant by $3.48$ Acc., $1.54$ AUROC, and $5.58$ F1 points.
\section{Conclusion}

We introduced MemeCF and MemePIVOT for cue-focused multimodal meme classification. MemeCF focuses on harm, hate, and sarcasm cases where the label depends on small but decisive textual, visual, or joint image--text cues, and provides evidence annotations describing the pivotal details behind each decision. MemePIVOT builds on this motivation by combining local word--patch grounding with global CLIP-based meme semantics, followed by evidential fusion to account for uncertainty and disagreement between local and global signals.

The experiments show that MemePIVOT is effective across the evaluated benchmarks. It achieves 88.04 accuracy, 92.92 AUROC, and 83.70 F1 on HarMeme, and 76.90 accuracy, 84.05 AUROC, and 76.98 F1 on PrideMM. On MemeCF, it obtains strong performance across all three subtasks: 71.38 accuracy, 78.20 AUROC, and 76.88 F1 for hate; 86.92 accuracy, 94.33 AUROC, and 87.92 F1 for harm; and 79.18 accuracy, 85.16 AUROC, and 82.18 F1 for sarcasm. The cross-dataset setting further indicates that the learned pivotal-evidence representation transfers well under domain shift, and the ablation study confirms the value of local grounding, evidential fusion, and optimal-transport alignment.

Future work can build directly on the evidence annotations in MemeCF by using them not only for dataset analysis, but also for explicit evidence-level supervision and evaluation. Another direction is to extend the cue-focused formulation to broader meme domains, languages, and cultural contexts, where pivotal cues may be more implicit or context-dependent. Finally, given the paper's finding that general-purpose multimodal LLMs remain costly and inconsistent in zero-shot harmful meme classification, future work can explore more efficient task-specific models that preserve fine-grained cue sensitivity while reducing inference cost.
\section*{Limitations}
Despite the complementary design of the local branch for subtle cue discovery and the global branch for holistic semantic understanding, the model remains fundamentally constrained by the frozen CLIP encoder used to produce its representations. In particular, culturally grounded humor, symbolism, slang, and historically situated references are only captured to the extent that they are already encoded in the pretrained backbone. As a result, the model may underperform on culturally sensitive memes whose harmful or sarcastic meaning depends on community-specific knowledge, implicit social context, or visual-textual references that are weakly represented in CLIP's pretraining distribution.

In addition, the evidential outputs should not be interpreted as a guarantee of calibrated reliability. Although the model produces Dirichlet-style evidence, belief, and uncertainty estimates, these quantities are still learned from the training distribution and inherit its biases and coverage gaps. Under distribution shift, unseen meme formats, unfamiliar cultural references, or ambiguous annotations, the model can still become overconfident despite producing uncertainty scores. Therefore, the uncertainty estimates are better viewed as a useful diagnostic signal rather than a substitute for calibration analysis, robustness testing, or human review.
\bibliography{references}

\section{Appendix}
\label{sec:appendix}
\subsection{Ablation Study: Evaluating the Performance of SOTA LLMs for Meme Classification}

A natural question one may wonder is whether a dedicated meme-classification model is still necessary in the 2026- given the rapid progress of large language models (LLMs) and their increasingly strong multimodal reasoning capabilities. To examine this, we perform an evaluation using state-of-the-art multimodal LLMs for harmful meme classification on the HarMeme's testset comprising of 354 images

For this evaluation, we select representative high-performing models from the MMMU-Pro~\citep{yue2025mmmupro} benchmark leaderboard, as visible in April 2026. From each model family, we choose the variant with the strongest reported performance. The evaluated models are: \texttt{anthropic/claude-opus-4.6}, \texttt{openai/gpt-5.5}, \texttt{google/gemini-3.1-pro-preview}, and \texttt{moonshotai/kimi-k2.6}~\citep{anthropic2026claudeopus46,openai2026gpt55,google2026gemini31pro,moonshot2026kimik26}. Following the MMMU-Pro evaluation setting, all models are evaluated under the high-reasoning configuration, except \texttt{openai/gpt-5.5}, for which the medium-reasoning setting is used because it achieves superior performance under the corresponding MMMU-Pro  benchmark configuration.

For each meme, we provide both the image and the associated text to the model. To ensure consistency across evaluations, we use a fixed random seed of 42. The below provided prompt is used-

\begin{quote}
\small
\texttt{DO NOT OUTPUT ANYTHING ELSE BESIDES 0 OR 1. OUTPUT ONLY FINAL ANSWER. Analyze the provided content. Determine if it qualifies as harmful according to the HarMeme definition, i.e., content that causes harm or has the potential to cause harm, including content potentially related to COVID-19 and targeting individuals, groups, or society. Output only `1' if harmful and `0' if not harmful based on the instructions.}
\end{quote}

Table~\ref{tab:model_evaluation_results} presents the performance of the evaluated LLMs on the HarMeme test set. Among the evaluated models, \texttt{anthropic/claude-opus-4.6} achieves the strongest performance, with an accuracy of 0.7881 and an AUROC of 0.7775. However, the remaining models perform substantially lower, with \texttt{openai/gpt-5.5}, \texttt{google/gemini-3.1-pro-preview}, and \texttt{moonshotai/kimi-k2.6} achieving accuracies of 0.6949, 0.6836, and 0.6176, respectively. The AUROC scores further indicate that several models struggle to reliably separate harmful from non-harmful memes, with \texttt{moonshotai/kimi-k2.6} performing only marginally above random discrimination.

These results suggest that even highly capable general-purpose multimodal LLMs do not fully solve harmful meme classification in a zero-shot setting. Meme classification remains challenging because it often requires understanding subtle interactions between visual content, overlaid text, cultural context, sarcasm, implicit targets, and potentially harmful intent. Therefore, relying solely on off-the-shelf LLMs may be insufficient for scalable harmful meme detection.

Another important limitation is computational cost. Even on a relatively small test set of only 354 memes, the models consume a large number of tokens. The total token usage ranges from 95,842 tokens for \texttt{anthropic/claude-opus-4.6} to 489,911 tokens for \texttt{google/gemini-3.1-pro-preview}. Such token consumption can lead to substantial inference costs when scaled to large meme datasets or real-world moderation pipelines involving thousands or millions of samples. In contrast, a dedicated task-specific model can provide a more efficient and cost-effective alternative while also being optimized directly for the harmful meme classification task.

\begin{table*}[t]
\centering
\small
\begin{tabular}{lrrrrr}
\hline
\textbf{Model} & \textbf{Accuracy} & \textbf{AUROC} & \textbf{Prompt Tokens} & \textbf{Completion Tokens} & \textbf{Total Tokens} \\
\hline
\texttt{anthropic/claude-opus-4.6} & 0.7881 & 0.7775 & 83,819 & 12,023 & 95,842 \\
\texttt{openai/gpt-5.5} & 0.6949 & 0.6073 & 69,461 & 40,732 & 110,193 \\
\texttt{google/gemini-3.1-pro-preview} & 0.6836 & 0.5837 & 344,968 & 144,943 & 489,911 \\
\texttt{moonshotai/kimi-k2.6} & 0.6176 & 0.5201 & 79,462 & 298,989 & 378,451 \\
\hline
\end{tabular}
\caption{Zero-shot harmful meme classification performance of state-of-the-art multimodal LLMs on the HarMeme test set.}
\label{tab:model_evaluation_results}
\end{table*}
\subsection{Implementation Details}

For all baseline models, we used the official released codebases or model checkpoints wherever available, and adapted only the dataset loading and classification heads required for our meme classification setting. Each model was trained on the corresponding training split, with the validation split used for hyperparameter tuning and best-checkpoint selection. Final performance was reported only on the held-out test split. To ensure fairness and reproducibility, experiments were run with three randomly chosen seeds. We report mean and standard deviation across seeds unless otherwise specified.

For non-VLM baselines, training was performed on an NVIDIA T4 GPU with 16GB VRAM. Vision-language model fine-tuning was performed on an NVIDIA A100 GPU with 40GB VRAM. For the VLM baselines, we used parameter-efficient fine-tuning wherever required to fit the models within the available hardware budget. As part of the experiments, for evaluation on PrideMM and HarMeme- we finetune LLaVA-1.5-7B, Qwen2.5-VL. For evaluation on MemeCF, we evaluate Gemma 3 Vision (12B), LLaVA-1.5-7B, Qwen3-VL-8B and Phi-4 Multimodal in a zero-shot setting

For LLaVA-1.5-7B, we followed the official LLaVA v1.5 task-LoRA fine-tuning recipe and adapted the instruction data to our binary meme classification setting. Specifically, dataset-specific JSON files were constructed for each task, and checkpoints and logs were maintained separately for each dataset. The meme image and the corresponding text transcription were included in the prompt, allowing the model to condition on both visual content and meme text. The best checkpoint was selected using validation performance before evaluation on the test set.

For Qwen2.5-VL, we used a dedicated multimodal fine-tuning wrapper. Each example was formatted as a conversation-style binary classification instance consisting of the meme image, a fixed classification instruction, and one of the allowed class labels as the target output. LoRA or QLoRA adapters were applied to the language-model attention and MLP layers, while the vision components were kept frozen by default. As with the other VLMs, validation macro-F1 was used for checkpoint selection.

During VLM evaluation, predictions were obtained by comparing the likelihood of the fixed candidate class labels rather than relying on unconstrained free-form generation. This produced deterministic label predictions and ensured that all VLMs were evaluated under the same binary classification protocol.
\subsection{Dataset Construction \& Statistics}

We construct MemeCF dataset using a semi-automated pipeline that combines LLM-as-a-judge with human validation. Using images from HarMeme, PrideMM, HMC, MultiBully, MMSD2, GuardHarMeme\cite{pramanick2021detecting,shah2024memeclip,kiela2020hateful,maity2022multitask,qin2023mmsd2,elamrany2025guardharmem}- we make use of GPT 5.5 model~\citep{openai2026gpt55} (having the highest performance on MMMU-Pro benchmark leaderboard at the time of dataset creation) for each meme to generate a task-specific binary label and structured counterfactual supervision. This supervision includes the modality of the decisive cue (\emph{text}, \emph{image}, or \emph{both}), the minimal evidence supporting the prediction, a minimal intervention expected to flip the label, and a short rationale explaining why the proposed intervention should change the class. To reduce annotation noise we follow a strict secondary validation using GPT 5.4 , Claude Opus 4.6 And Gemini Flash 3 model~\citep{openai2026gpt54,anthropic2026claudeopus46,google2025gemini3flash} as the judges-  retaining only outputs with valid binary labels and confidence scores of at least $0.9$. The retained pool is then manually reviewed to remove noisy, ambiguous, low-quality, or otherwise unreliable examples. Finally, accepted examples are assigned to train, validation, and test partitions \\\begin{figure*}
    \centering
    \includegraphics[width=\textwidth]{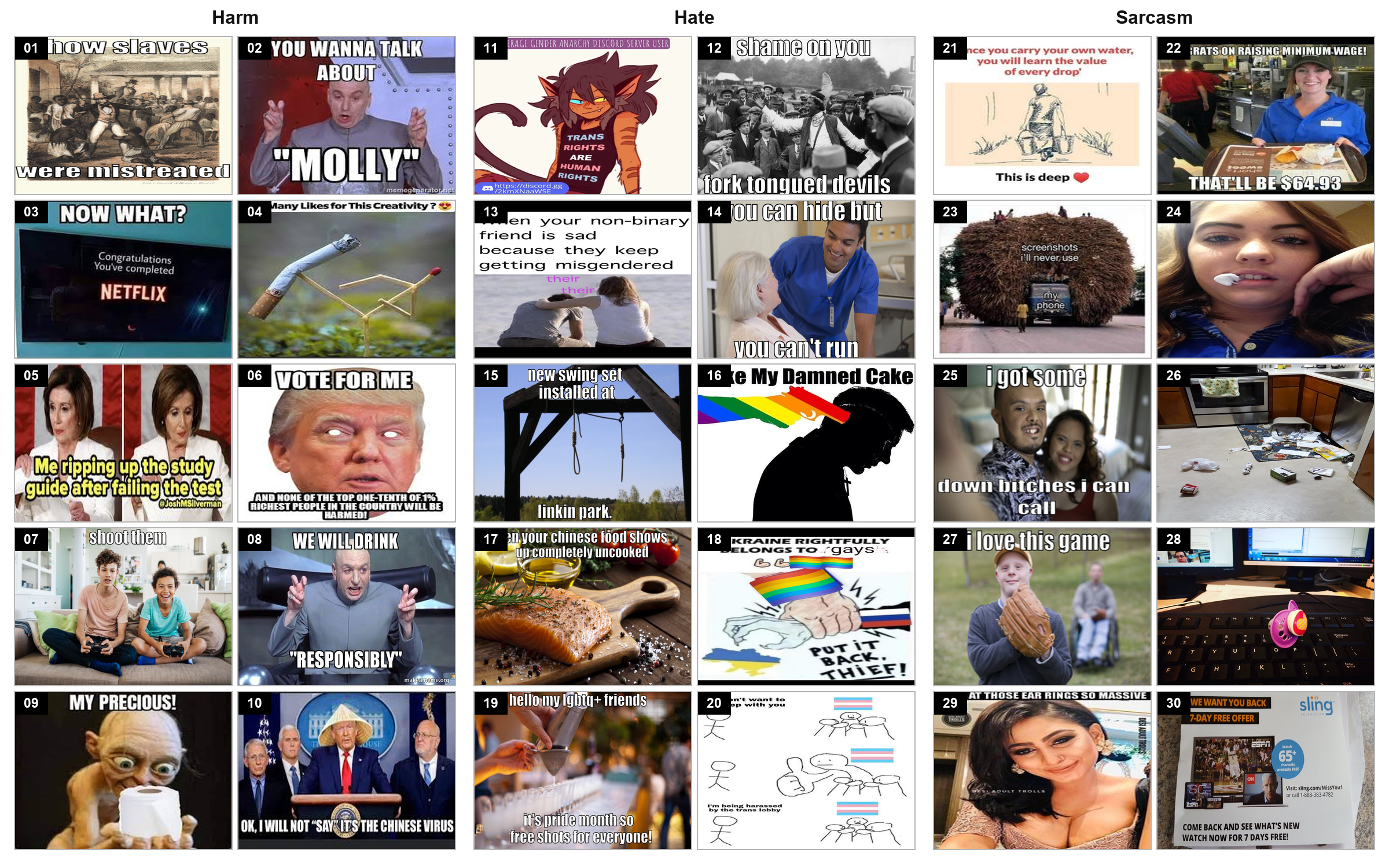}
    \caption{Thirty representative non-trivial MemeCF examples across harm, hate, and sarcasm. Within each task pair, the label-0 example appears on the left and the label-1 example on the right; the numbered examples correspond to the annotations in Table \ref{tab:representative-examples}.}
    \label{fig:representative_examples}
\end{figure*}

The dataset contains three task-specific subsets: \emph{harm}, \emph{hate}, and \emph{sarcasm}. The harm subset is derived from GuardHarMem and HarMeme, the hate subset from PrideMM and HMC, and the sarcasm subset from MMSD2 and MultiBully. Across the three tasks, the benchmark contains 9,895 memes in total, comprising 7,421 training memes, 990 validation memes, and 1,484 test memes, corresponding to an approximately $75/10/15$ split.\\
The dataset annotations contain the meme image, associated OCR or caption text, task label, data split, source information, and counterfactual supervision. Label semantics are task-specific: for harm, $0$ denotes non-harmful and $1$ denotes harmful; for hate, $0$ denotes non-hateful and $1$ denotes hateful; and for sarcasm, $0$ denotes non-sarcastic and $1$ denotes sarcastic. Counterfactual supervision specifies the decisive cue, a proposed intervention that would flip the label, and a rationale explaining the expected change.
The harm subset contains 3,018 memes, with 1,962 from GuardHarMem ($65.0\%$) and 1,056 from HarMeme ($35.0\%$). Its label-0/label-1 distribution is 1,397/1,621. The subset is partitioned into 2,263 training, 302 validation, and 453 test memes. Its counterfactual cues comprise 2,374 text-based, 411 image-based, and 233 multimodal examples. The mean text length is 74.0 characters, with a median of 62 characters.\\

The hate subset contains 3,289 memes, sourced from 1,735 PrideMM memes ($52.8\%$) and 1,554 HMC memes ($47.2\%$). Its label-0/label-1 distribution is 1,500/1,789. The subset contains 2,467 training, 329 validation, and 493 test memes. Its counterfactual cues comprise 2,783 text-based, 377 image-based, and 129 multimodal examples. The mean text length is 108.0 characters, with a median of 73 characters.\\

The sarcasm subset contains 3,588 memes, including 2,029 from MMSD2 ($56.5\%$) and 1,559 from MultiBully ($43.5\%$). Its label-0/label-1 distribution is 1,639/1,949. The subset is divided into 2,691 training, 359 validation, and 538 test memes. Its counterfactual cues comprise 2,338 text-based, 1,100 image-based, and 150 multimodal examples. The mean text length is 58.5 characters, with a median of 53 characters. MemeCF thereby provides a standardized format and supervision style across all three tasks while preserving source diversity and task-specific label semantics. Across the full collection, text-based counterfactuals are the most common, accounting for 7,495 memes, followed by 1,888 image-based memes and 512 multimodal memes. This design makes the benchmark suitable not only for binary meme classification, but also for counterfactual reasoning, evidence attribution, and multimodal analysis. Figure \ref{fig:representative_examples} shows 30 non-trivial examples and Table \ref{tab:representative-examples} provides the corresponding annotations.

\onecolumn

\begingroup
\scriptsize
\setlength{\tabcolsep}{2pt}
\renewcommand{\arraystretch}{1.05}
\setlength{\LTpre}{0pt}
\setlength{\LTpost}{0pt}
\setlength{\LTcapwidth}{\textwidth}
\setlength{\LTleft}{0pt}
\setlength{\LTright}{0pt}
\setlength{\emergencystretch}{3em}

\begin{longtable}{@{}
>{\Centering\arraybackslash}p{0.60cm}
>{\RaggedRight\arraybackslash}p{1.85cm}
>{\RaggedRight\arraybackslash}p{0.40\textwidth}
>{\RaggedRight\arraybackslash}p{\dimexpr\textwidth-2.45cm-0.40\textwidth-6\tabcolsep\relax}
@{}}
\caption{Thirty representative non-trivial MemeCF examples and their cue-focused annotations.} \label{tab:representative-examples}\\

\toprule
No. & Task / Type / Label & Text / Evidence & Flip / Rationale \\
\midrule
\endfirsthead

\toprule
No. & Task / Type / Label & Text / Evidence & Flip / Rationale \\
\midrule
\endhead

01 & \RepMetaCell{Harm}{text}{0} & \RepSourceCell{how slaves were mistreated}{the word 'mistreated'} & \RepFlipCell{change text word 'mistreated' to 'disciplined'}{The word 'mistreated' is the dominant local cue that makes the meme a condemnatory historical statement rather than hateful content. Replacing it with 'disciplined' reframes the same slavery scene as justified punishment, substantially changing the meme's core meaning and making it harmful toward a protected group.} \\

02 & \RepMetaCell{Harm}{text}{1} & \RepSourceCell{you wanna talk about "molly"}{the word 'molly'} & \RepFlipCell{change the text word 'MOLLY' to 'MUSIC'}{The quoted word 'MOLLY' is the main label-bearing cue because it is the only explicit content tied to a drug reference; the image is just a generic reaction pose. Replacing that single word with a harmless topic like 'MUSIC' keeps the meme coherent but removes the harmful drug connotation, so the label flips to non-harmful.} \\

03 & \RepMetaCell{Harm}{text}{0} & \RepSourceCell{NOW WHAT? Congratulations You've completed NETFLIX}{the word 'NETFLIX'} & \RepFlipCell{change text from 'NETFLIX' to 'Jews'}{The single word 'NETFLIX' is the dominant local cue that makes the meme a harmless joke about binge-watching a streaming service. Replacing that noun with 'Jews' changes the core meaning from finishing a TV catalog to 'completing' a protected group, which reads as violent/eliminationist and flips the meme to harmful.} \\

04 & \RepMetaCell{Harm}{image}{1} & \RepSourceCell{how many likes for this creativity?}{a cigarette being kicked by a matchstick figure} & \RepFlipCell{Replace the cigarette on the left with a non-smoking object, such as a pencil or lollipop, while keeping the rest of the image unchanged.}{The cigarette is the main local evidence for the harmful label because it directly signals smoking/tobacco use, which makes the meme harmful. Swapping it for a benign object preserves the ``creative craft'' idea but removes the harmful association, flipping the label to non-harmful.} \\

05 & \RepMetaCell{Harm}{text}{0} & \RepSourceCell{Meripping up the study guide after failing the test loshMSIlverman}{study guide} & \RepFlipCell{change text from 'study guide' to 'Quran'}{The phrase 'study guide' is the compact object being ripped up and is the main cue that the meme is a harmless school-related joke. Replacing it with 'Quran' changes the act into destruction of a religious text, shifting the meaning from frustration about a failed test to religiously hostile desecration, which would flip the label to harmful.} \\

06 & \RepMetaCell{Harm}{image}{1} & \RepSourceCell{VOTE FOR ME AND HONE OF THE TOP ONE-TENTH OF 1\% RICHEST PEOPLE IN THE COUNTRYWILL BE HARMEDI}{The eyes of Donald Trump are edited to be solid white, giving a demonic or soulless appearance.} & \RepFlipCell{change Trump's blank white eyes to normal human eyes with visible pupils/irises}{The strongest local evidence for harmfulness is the eye region: the blank white eyes make Trump look monstrous/possessed and turn the meme from ordinary political criticism into a hostile, dehumanizing depiction. Replacing just that compact visual attribute with normal eyes keeps the meme coherent but substantially changes its tone and meaning, making it a non-harmful political joke rather than a harmful attack.} \\

07 & \RepMetaCell{Harm}{image}{0} & \RepSourceCell{shoot them}{the boys are holding video game controllers} & \RepFlipCell{change the pair of video game controllers in the children's hands to real guns}{The dominant local evidence making the meme non-harmful is the compact visual cue that the children are playing a video game. The pair of controllers is the smallest semantically complete image region that carries most of that signal. Replacing them with real guns changes the core meaning of 'shoot them' from a likely in-game action to a real-world call for violence, which flips the label to harmful.} \\

08 & \RepMetaCell{Harm}{both}{1} & \RepSourceCell{we will drink "responsibly"}{Text: quotation marks around the word 'responsibly'; Visual: Dr. Evil's finger air quotes gesture} & \RepFlipCell{remove the quotation marks around 'responsibly' and change the image's air-quote hand gesture to a neutral/non-quoting pose}{The dominant harmfulness cue is the sarcastic scare-quote concept around 'responsibly', expressed both in the text quotation marks and the character's air-quote gesture. That local unit turns the statement into mockery of responsible drinking, implying irresponsible alcohol use. Replacing both cues with sincere, non-sarcastic versions changes the core meaning to genuinely drinking responsibly, which flips the meme to non-harmful.} \\

09 & \RepMetaCell{Harm}{image}{0} & \RepSourceCell{MY PRECIOUS!}{a roll of toilet paper} & \RepFlipCell{change the toilet paper roll in Gollum's hands to a swastika symbol}{The toilet paper roll is the smallest semantically complete visual unit that makes the meme a harmless joke about obsessing over an everyday item. Replacing that single object with a swastika substantially changes the core meaning from benign fixation to affectionate valuation of a hate symbol, which would make the meme harmful.} \\

10 & \RepMetaCell{Harm}{both}{1} & \RepSourceCell{OUSE OK, I WILL NOT "SAY ITS THE CHINESE VIRUS}{Text: The specific phrase 'CHINESE VIRUS' in the caption.; Visual: The edited conical straw hat (non-la) on Donald Trump's head.} & \RepFlipCell{change text from 'CHINESE' to 'COVID-19' and change the image conical Asian hat on Trump to his normal hair}{The harmful meaning is driven by the combined China/Asian reference: the text uses the stigmatizing phrase 'Chinese virus' while the image reinforces it with a stereotyped conical Asian hat. Changing only the word or only the hat would leave the other dominant cue intact, but replacing both removes the xenophobic attribution and makes the meme a generic joke about avoiding a phrase, flipping it to non-harmful.} \\

11 & \RepMetaCell{Hate}{text}{0} & \RepSourceCell{average gender anarchy discord server user trans rights are human rights 2kmxnaawse}{the phrase 'trans rights are human rights'} & \RepFlipCell{change the shirt text from 'TRANS RIGHTS ARE HUMAN RIGHTS' to 'TRANS RIGHTS AREN'T HUMAN RIGHTS'}{The shirt slogan is the dominant label-bearing cue; it directly states the meme's stance. Replacing it with an anti-trans version flips the core meaning from supportive/non-hateful to hateful while keeping the meme coherent.} \\

12 & \RepMetaCell{Hate}{text}{1} & \RepSourceCell{shame on you fork tongued devils}{fork tongued devils} & \RepFlipCell{change text from 'fork tongued' to 'lying'}{The two-word phrase 'fork tongued' is the smallest dominant label-bearing unit. It invokes a racialized stereotype tied to Native Americans, while the rest of the caption is mainly generic condemnation. Replacing it with 'lying' preserves a coherent insult but removes the stereotype-driven hateful meaning, which should flip the meme to non-hateful.} \\

13 & \RepMetaCell{Hate}{text}{0} & \RepSourceCell{when your non binary friend is sad because they keep getting misgendered their their}{the word 'their' used as a pun for 'there'} & \RepFlipCell{change the overlaid phrase from 'their their' to 'it it'}{The small overlaid phrase 'their their' is the meme's punchline and main label-bearing cue: it reframes the comforting phrase 'there there' with a non-binary pronoun, making the meme sympathetic rather than hostile. Replacing that phrase with 'it it' changes the core meaning from supportive wordplay to dehumanizing misgendering of a non-binary person, which would make the meme hateful.} \\

14 & \RepMetaCell{Hate}{image}{1} & \RepSourceCell{you can hide but you can't run}{the presence of an elderly woman who appears to have limited mobility} & \RepFlipCell{change the elderly seated woman to a clearly younger adult woman in the same pose and setting}{The dominant label-bearing cue is the image's depiction of an elderly target. The text is generic on its own, but paired with an obviously old woman it becomes an age-targeted taunt about her inability to run, which makes the meme hateful. Replacing the elderly woman with a younger adult removes the protected-age targeting while keeping the scene coherent, flipping the meme to non-hateful.} \\

15 & \RepMetaCell{Hate}{text}{0} & \RepSourceCell{new swing set installed at linkin park}{linkin park} & \RepFlipCell{change text from 'linkin park' to 'the mosque'}{The phrase 'linkin park' is the key local cue that makes the noose image read as a dark joke about the band Linkin Park/Chester Bennington rather than as hostility toward a protected group. Replacing only that phrase with 'the mosque' keeps the meme coherent but changes the target to a Muslim religious site, so the hanging image becomes an implied anti-Muslim threat and the label flips to hateful.} \\

16 & \RepMetaCell{Hate}{image}{1} & \RepSourceCell{bake my damned cake}{the rainbow colors (LGBTQ+ flag) applied to the gun silhouette} & \RepFlipCell{change the rainbow pride-colored stripes hitting the silhouette's head to a plain neutral gray paint streak}{The rainbow pride-colored visual cue is the dominant local evidence identifying the target as LGBTQ people and framing them as aggressively demanding service. The text alone is just an angry demand about cake; without the pride cue, the meme becomes a generic complaint rather than anti-LGBTQ. Replacing the rainbow with a neutral paint streak removes the protected-class targeting and flips the meme to non-hateful.} \\

17 & \RepMetaCell{Hate}{image}{0} & \RepSourceCell{when your chinese food shows up completely uncooked}{a piece of raw salmon} & \RepFlipCell{change the raw salmon fillet on the cutting board to a clearly recognizable dead dog on the cutting board}{The salmon fillet is the dominant local visual cue that makes the meme a benign joke about raw food. Replacing that single food item with a recognizable dog would substantially change the meaning from 'uncooked meal' to the racist stereotype that Chinese food is dog meat, which would make the meme hateful toward Chinese people.} \\

18 & \RepMetaCell{Hate}{both}{1} & \RepSourceCell{ukraine rightfully belongs to gays bb put it back thief}{Text: The word 'gays' in the top caption; Visual: The rainbow flag armbands and icons overlaid on the muscular arm} & \RepFlipCell{change text from 'gays' to 'ukrainians' and change the image rainbow flag on the fist to a Ukrainian flag}{The local evidence identifying the aggressor as gay/LGBTQ is carried by the word 'gays' together with the rainbow flag on the fist. This protected-class cue is the dominant reason the meme is hateful, because it frames gays as thieves who wrongfully take Ukraine. Replacing that paired cue with 'ukrainians' and a Ukrainian flag changes the target from a protected group to a national self-ownership message, making the meme non-hateful.} \\

19 & \RepMetaCell{Hate}{image}{0} & \RepSourceCell{hello my lgbtq+ friends it's pride month so free shots for everyone!}{the liquid being poured from the shaker} & \RepFlipCell{change the local drink-shot region (the row of shot glasses being served/poured) to a handgun being aimed or fired instead of drink shots}{The shot-glass serving region is the smallest dominant cue that disambiguates 'shots' as alcohol, which makes the meme a pride-month celebration and therefore non-hateful. Replacing that compact visual unit with a gun/firearm scene would reinterpret 'free shots' as gunshots directed at the mentioned LGBTQ+ group, substantially changing the meme's meaning and flipping the label to hateful.} \\

20 & \RepMetaCell{Hate}{both}{1} & \RepSourceCell{i don't want to sleep with you i'm being harassed by the trans lobby 11}{Text: The phrase 'trans lobby' in the bottom text.; Visual: The transgender pride flag appearing above the group of stick figures.} & \RepFlipCell{change text from 'trans lobby' to 'dating app lobby' and change the trans pride flags to generic dating app icons}{The dominant hateful signal is the local identification of the harassing group as trans, reinforced by the trans pride flags. Replacing that protected-group cue with a non-protected generic dating-app cue changes the meme from portraying trans people as sexual harassers to a non-hateful joke about app-driven harassment, flipping the label.} \\

21 & \RepMetaCell{Sarcasm}{text}{0} & \RepSourceCell{Once you carry your own water,you will learn the value of every drop'This is a deep}{the heart emoji at the bottom} & \RepFlipCell{change the heart emoji in 'This is deep \(\heartsuit\)' to a rolling-eyes emoji}{The heart emoji is the smallest local tone cue making the bottom caption read as sincere appreciation of the inspirational quote. Replacing it with a rolling-eyes emoji turns 'This is deep' into a mocking reaction, substantially changing the meme from earnest motivational content to sarcastic dismissal.} \\

22 & \RepMetaCell{Sarcasm}{text}{1} & \RepSourceCell{wouldn 't it be awesome if everyone could make \$ 75 per hour ?}{\$64.93} & \RepFlipCell{Change the text "\$64.93" to "\$6.49".}{The exaggerated price is the main sarcastic punchline. Replacing it with a normal fast-food price removes the irony, making the meme read as a straightforward transaction rather than a sarcastic complaint.} \\

23 & \RepMetaCell{Sarcasm}{text}{0} & \RepSourceCell{Screenshots I'll never use my phone}{the word 'never' in the phrase 'screenshots i'll never use'} & \RepFlipCell{change text from 'never' to 'always'}{The word 'never' is the smallest dominant text unit that makes the meme a literal, non-sarcastic complaint about storing many useless screenshots. Replacing it with 'always' creates a clear contradiction between the enormous pile of screenshots and the claim that they will all be used, making the meme sarcastic.} \\

24 & \RepMetaCell{Sarcasm}{image}{1} & \RepSourceCell{man i love the dentist .}{the white cotton ball/gauze tucked into the side of the woman's mouth} & \RepFlipCell{replace the cotton/gauze wad sticking out of her mouth with a normal closed mouth or relaxed smile}{The protruding dental gauze is the dominant local cue that signals an unpleasant just-left-the-dentist situation, which directly contradicts the positive text 'man i love the dentist .' and creates the sarcasm. Removing that single visual cue makes the selfie no longer depict obvious dental discomfort, so the text reads as straightforward rather than sarcastic.} \\

25 & \RepMetaCell{Sarcasm}{text}{0} & \RepSourceCell{i got some down bitches i can call}{the word 'down'} & \RepFlipCell{change the word 'down' to 'fine'}{The single word 'down' is the dominant local cue because it makes the caption align with the pictured woman as a Down-syndrome joke rather than an ironic boast. That alignment is what keeps the meme non-sarcastic. Replacing 'down' with 'fine' changes the core meaning to bragging about attractive women, which the image then undercuts, creating sarcasm.} \\

26 & \RepMetaCell{Sarcasm}{image}{1} & \RepSourceCell{love it when my mothers dogs are here .}{the knocked-over trash can and scattered garbage in the kitchen} & \RepFlipCell{change the overturned trash can and scattered garbage on the kitchen floor to a clean, undisturbed floor with the trash can upright}{The dominant evidence for sarcasm is the local visual mess caused by the dogs: the overturned bin and spilled trash directly contradict the positive text 'love it when my mothers dogs are here.' This compact visual region carries most of the sarcastic signal. Replacing it with a clean floor and upright bin changes the implied behavior of the dogs from destructive to harmless, making the text read as sincere rather than sarcastic.} \\

27 & \RepMetaCell{Sarcasm}{image}{0} & \RepSourceCell{i love this game}{the smiling, happy facial expression of the boy with Down syndrome} & \RepFlipCell{change the boy's calm/happy face to a crying or visibly upset face}{The boy's positive facial expression is the main local visual cue that makes the caption 'i love this game' read sincerely. Replacing only his face with an upset expression creates a clear contradiction between the enthusiastic text and the negative reaction, making the meme read sarcastically.} \\

28 & \RepMetaCell{Sarcasm}{both}{1} & \RepSourceCell{ways to impress higher ups : reach for a pen and pull a paci out of your pocket .}{Text: the word 'paci' in the text; Visual: the purple pacifier sitting on the keyboard} & \RepFlipCell{change text from 'paci' to 'pen' and change the image object from a pacifier to a pen}{The pacifier is the dominant local cue carrying the sarcasm: it makes the supposed advice for impressing higher-ups obviously childish and counterproductive. Replacing that single entity with a pen changes the core meaning from embarrassing yourself to appearing prepared/professional, which makes the meme read as non-sarcastic advice.} \\

29 & \RepMetaCell{Sarcasm}{image}{0} & \RepSourceCell{look at those ear rings so massive}{the size of the earrings on the woman's ears} & \RepFlipCell{change the woman's large flower-shaped earrings to tiny plain stud earrings}{The dominant evidence for the current non-sarcastic label is the visible oversized earrings, which make the caption's literal claim coherent. The earrings are a compact, semantically complete visual unit. Replacing them with tiny studs would make the text 'look at those ear rings so massive' clearly contradict the image, substantially changing the meme's meaning and flipping it to sarcastic.} \\

30 & \RepMetaCell{Sarcasm}{both}{1} & \RepSourceCell{oh wow ! thank u so much <user> you must really want me back . <num> whole days ? ! what an offer ! icymi}{Text: The phrase '7 whole days'; Visual: The text '7-DAY' on the physical mailer} & \RepFlipCell{change text from '<num> whole days' to '90 whole days' and change the image offer duration from '7-DAY/7 DAYS FREE' to '90-DAY/90 DAYS FREE'}{The dominant local evidence is the short 7-day free-offer duration, which appears in both the meme text and the flyer. The sarcasm comes from treating a very small win-back offer as if it were generous. Replacing the duration with a much more generous 90-day offer makes the praise coherent and sincere, flipping the meme to non-sarcastic.} \\

\bottomrule
\end{longtable}

\endgroup

\clearpage
\twocolumn
\end{document}